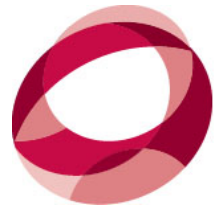

**Privacy-Preserving Data Analysis for the Federal Statistical Agencies**

January 2017

*John Abowd, Lorenzo Alvisi, Cynthia Dwork, Sampath Kannan, Ashwin Machanavajjhala, and Jerome Reiter*

Government statistical agencies collect enormously valuable data on the nation's population and business activities. Wide access to these data enables evidence-based policy making, supports new research that improves society, facilitates training for students in data science, and provides resources for the public to better understand and participate in their society. These data also affect the private sector. For example, the Employment Situation in the United States, published by the Bureau of Labor Statistics, moves markets. Nonetheless, government agencies are under increasing pressure to limit access to data because of a growing understanding of the threats to data privacy and confidentiality.

"De-identification" — stripping obvious identifiers like names, addresses, and identification numbers — has been found inadequate in the face of modern computational and informational resources (Sweeney 2007; Narayanan and Shmatikov 2006; Narayanan and Shmatikov 2010; Sweeney 2013; see also the report of the President's Council of Advisors on Science and Technology 2014).

Unfortunately, the problem extends even to the release of aggregate data statistics (Dinur and Nissim 2003; Dwork, McSherry, and Talwar 2007; Homer et al. 2008; Kasiviswanathan, Rudelson, Smith, and Ullman 2010; De 2012; Kasiviswanathan, Rudelson, and Smith, 2013; Muthukrishnan and Nikolov 2012; Dwork et al 2015). This counter-intuitive phenomenon has come to be known as the Fundamental Law of Information Recovery. It says that overly accurate estimates of too many statistics can completely destroy privacy. One may think of this as death by a thousand cuts. Every statistic computed from a data set leaks a small amount of information about each member of the data set – a tiny cut. This is true even if the exact value of the statistic is distorted a bit in order to preserve privacy. But while each statistical release is an almost harmless little cut in terms of privacy risk for any individual, the cumulative effect can be to completely compromise the privacy of some individuals. Since statistical releases serve useful purposes, statistical bureaus must strategically choose which cuts to inflict, while protecting privacy overall. The problem is especially acute given the level of detail at which these bureaus collect and release data periodically. The Census Bureau, for example, plans to publish several billion statistical summaries from the 2020 Census of Population and Housing.

The Fundamental Law of Information Recovery has troubling implications for the publication of large numbers of statistics by a statistical agency: it says that the confidential data may be vulnerable to database reconstruction attacks based entirely on the data published by the agency itself. Left unattended, such risks threaten to undermine, or even eliminate, the societal benefits inherent in the rich data collected by the nation's statistical agencies. The most pressing immediate problem for any statistical agency is how to modernize its disclosure limitation methods in light of the Fundamental Law.

The technical challenge to the federal statistical agencies is to quantify the inferential disclosure (privacy loss) and mathematically bound, or control, it for the complete publication plan for any set of confidential data inputs; that is, quantify the sum of many "tiny" privacy cuts. *Differential privacy*, a definition of privacy and a collection of supporting algorithmic techniques tailored for privacy-preserving statistical analysis of large datasets, provides such a measure of privacy loss, together with the tools to analyze and control cumulative loss accruing over multiple publications (Dwork, McSherry, Nissim, and Smith, 2006; Dwork, 2006). In principle, using this technology, an agency can set a privacy-loss "budget", measured in the units supplied by differential privacy, and ensure that its entire publication plan stays within budget. It can prioritize certain statistics, using a larger share of the budget and having better accuracy, and lower the priority on others, resulting in less privacy budget consumption but worse accuracy for these lower priority products.

This apparent "match made in heaven" between differential privacy and the needs of a statistical agency holds great promise but requires great effort. There are two principal areas of difficulty:

1) The Fundamental Law of Information Recovery can no more be circumvented than can the laws of physics. Consider the billions of statistical summaries planned for the 2020 Census of Population and Housing. If Census imposes a binding privacy-loss ceiling on these publications, how will that budget be distributed across the competing interests of accuracy for legislative redistricting (detail at small geographic levels--census blocks) versus data on small subpopulations (detail on many ethnic and racial minorities)? No statistical agency has ever had to examine its publication policies under the glare of an explicit privacy-loss budget. Yet, that is precisely what is demanded by acknowledging the Fundamental Law.
2) Differential privacy is a young field. The literature is silent on crucial preprocessing steps, such as imputation of missing fields and other aspects of data cleaning. Algorithmic challenges specific to the statistical agencies are just now being formulated, and their solution will be the subject of doctoral dissertations as yet unwritten. Better techniques are needed to communicate the error experienced by analysts. Expertise in differential privacy – introduced a scant decade ago -- is geographically scattered. Recently announced adoption of the approach by Apple and Google will draw talent away from the public and research sectors.  To use formal privacy-preserving disclosure limitation methods for the 2020 Census, the methods must be developed, tested, and subjected to scientific scrutiny before the conclusion of the 2018 End-to-end test (roughly, February 2019).

A national Virtual Center for Differential Privacy (V-CDP) could harness talent to address the privacy-preserving data analysis challenges of the federal statistical system.  The techniques and artifacts developed would be useful to statistical agencies, as well as industry, but it is the Census Bureau that has recognized the need for formal privacy guarantees and is reaching out for help from the research community.

**Structure**

A V-CDP could build an infrastructure and a set of tools and techniques for safeguarding the privacy of individuals and groups, while allowing for valuable analyses of census data.  The driving idea for this V-CDP is that major new algorithmic breakthroughs are needed if we want to be able to return results that are sufficiently accurate for their statistical purposes, while staying within a meaningful privacy budget[1].  Current approaches are too *ad hoc* and frequently do not achieve anything close to the best trade-offs between accuracy and privacy. Researchers in computer science and statistics are key people needed to carry out this agenda.  The highest priorities for the V-CDP would be the design of techniques needed for analyzing data from the 2020 Census of Population and Housing, the 2017 Economic Census, and the annual American Community Surveys going forward.

The Virtual Center for Differential Privacy would convene computer scientists, statisticians, mathematicians, and economists in sustained interactions and collaborations and will provide many resources and modes of collaboration to its participants:
1) It would be a repository for public decennial census data and annual data from American Community Surveys. There will be enough computing power at the center for members to remotely run software on data at the center.
2) It would serve as a repository for papers and software developed.
3) It would have the technology to act as a hub for state-of-the-art videoconferencing between members.

---

[1] Differential privacy offers a measure of privacy loss and tools for understanding how this loss accumulates over multiple computations.  This makes it possible to specify a "privacy loss budget" and to ensure that the cumulative privacy loss stays within the budget.

The V-CDP would have a managing board consisting of a representative of the Census Bureau, together with a computer scientist and a statistician with affiliations outside the census bureau.

**Casting A Wider Net**

To bring modern privacy techniques into the federal statistical system requires the knowledge, experience, and perspectives of multiple disciplines. These include not only computer scientists, statisticians, and applied mathematicians--all of whom have contributed to the development of privacy-preserving methods and bring essential and complementary skills to the V-CDP--but also social and behavioral scientists, legal scholars, and government employees, to name a few. Social and behavioral scientists focus the algorithm development. They help answer the question: With a limited privacy budget, where should the published data be most accurate? Legal scholars are needed to ensure that products satisfy regulatory and ethical obligations to protect privacy while also meeting reporting requirements typical for federal statistical data. We look for opportunities for feedback loops: Advances in privacy-preserving methods could point the way to better laws. For example, most federal laws focus on risks of re-identification in data, but a focus on inferential disclosures would result in both greater protection and more accuracy. Finally, active participants from the federal statistical system are needed to steer algorithm development toward the practical problems of interest, so that methods are implemented to create genuinely useful data products. Experienced producers of federal statistics bring a wealth of "on-the-ground" experience with these data products to the V-CDP: How the data are collected; how missing and erroneous values arise and are handled; what secondary users expect the federal agencies to share. In short, developing and implementing formally private algorithms to create safe and useful federal data products requires interdisciplinary collaborations on the scope of a V-CDP.

**Education**

Education is an important tool for the sustained adoption of differentially private and other provably private methodologies for releasing/accessing sensitive federal datasets. At least four communities will be stakeholders in an ecosystem that allows safe access to sensitive data:
1. Inventors/Designers of safe release practices: These might include computer scientists, statisticians and applied mathematicians who will design the next generation of algorithms for safe release of federal data while ensuring that the data remain useful to tackle the novel challenges that arise from federal datasets (high dimensionality, complex data types, graphs, longitudinal data, etc.) Education for this community would include a thorough understanding of the basics of differential privacy and other formal privacy notions, and the algorithmic advances thus far in this field. There are courses in academic institutions (e.g., UPenn, Duke, Harvard), tutorials (e.g. VLDB 16 tutorial) and books (Dwork-Roth) that cover the algorithmic aspects of differential privacy and other formal privacy notions, as well as techniques for releasing and analyzing sensitive data, which could be collated into a MOOC style course. These courses should also expose inventors to the typical uses of data (regression, log linear modeling, imputation, data cleaning, etc) so that end-to-end provably private methodologies can be designed.
2. Users of released data: These typically include economists, social scientists and others who use data released by federal agencies on a regular basis. While formal privacy techniques can release synthetic data that look like real data, an astute data user must know that such data can only be trusted for certain aggregate analyses and not for other more fine-grained analyses. Moreover, an understanding of how data are perturbed can help design new analysis methods that can extract more utility from the released data. Education for this community might include online courses, like the one at Cornell University on the use of synthetic data (Abowd and Vilhuber), and basics of private data release and analysis.
3. Deployers of release practices: Due to the volume and variety of data released by federal agencies, we envision that not all deployers of release methodologies are inventors of privacy algorithms. While deployers need training on the basics of differential privacy (privacy budgets, building blocks, etc) they also need a good understanding of the space of solutions that are applicable for their data and analysis needs, as well as a good understanding of the privacy/utility tradeoffs on benchmark datasets. In addition to lectures/tutorials on privacy, expert-vetted open source implementations of privacy algorithms (e.g., PSI) and visualizations of their performance on benchmark datasets (e.g., DPComp.org) would be great resources for deployers.
4. Policy makers around what constitutes a safe release: Finally, there has to be a dialogue between policy makers and inventors of privacy definitions and algorithms. This dialogue is essential to help develop the next generation of regulations that go beyond simple measures like re-identification and safe harbor. We envision developing courses jointly with, say, privacy law scholars and computer scientists wherein mathematical underpinnings of formal privacy notions (e.g., the fundamental law of information disclosure, why Bayesian risk factors are the right measure for privacy loss and their interpretation, etc.) are communicated to policy makers, while current policy and regulations are communicated with inventors of mathematical privacy definitions.

# Citations

*This material is based upon work supported by the National Science Foundation under Grant No. (1136993). Any opinions, findings, and conclusions or recommendations expressed in this material are those of the author(s) and do not necessarily reflect the views of the National Science Foundation.*